\documentstyle[sprocl,epsfig]{article}

\def\Journal#1#2#3#4{{#1} {\bf #2}, #3 (#4)}

\def\NPB{{\em Nucl. Phys.} B}
\def\PLB{{\em Phys. Lett.}  B}
\def\PRL{\em Phys. Rev. Lett.}
\def\PRD{{\em Phys. Rev.} D}

\def\be{\begin{equation}}
\def\ee{\end{equation}}
\def\bea{\begin{eqnarray}}
\def\eea{\end{eqnarray}}

\begin{document}


\setcounter{page}{0}
\pagestyle{empty}

\begin{flushright}
HIP-2000-56/TH 
\end{flushright}
\vspace*{26mm}

\title{THE EFFECT OF PRIMORDIAL TEMPERATURE FLUCTUATIONS 
       ON THE QCD TRANSITION\protect\footnotemark[1] \vspace*{8mm} }
\footnotetext[1]{
 To appear in the Proceedings of Strong and Electroweak Matter 2000
 (SEWM2000), ${\mbox{\rm Marseilles}}$, France, 14--17 June 2000.}

\author{J. IGNATIUS \vspace*{1mm} }

\address{Department of Physics,
         P.O. Box 9, FIN-00014 University of Helsinki, Finland\\
E-mail: janne.ignatius@iki.fi  \vspace*{3mm} }

\author{DOMINIK J. SCHWARZ \vspace*{1mm} }

\address{Institut f\"ur Theoretische Physik, 
         TU Wien,\\ 
         Wiedner Hauptstra\ss e 8 -- 10, A-1040 Wien, Austria\\ 
E-mail: dschwarz@hep.itp.tuwien.ac.at \vspace*{8mm}  }


\maketitle\abstracts{
We 
analyse                 
a new mechanism for the cosmological QCD
first-order             
phase transition:
inhomogeneous nucleation. The primordial temperature fluctuations
are larger than the tiny temperature 
interval, in which bubbles would form in the standard picture of homogeneous 
nucleation. Thus the bubbles nucleate at cold spots. We find the typical 
distance between bubble centers to be a few meters. This exceeds 
the estimates from homogeneous nucleation by two orders of magnitude. 
The resulting baryon inhomogeneities may affect primordial nucleosynthesis. 
}

\clearpage

\pagestyle{plain}

First-order phase transitions normally proceed via nucleation of bubbles 
of the new phase. When the temperature is spatially uniform and no 
impurities are present, the mechanism is homogeneous nucleation. 
%
We denote 
by $v_{\rm heat}$ the effective speed by which released latent heat 
propagates in sufficient amounts to shut down nucleations.
%
The mean distance between nucleation centers, measured immediately after 
the transition completed, is
\begin{equation}
  d_{\rm nuc,hom} = 2 v_{\rm heat} \Delta t_{\rm nuc}
  \label{dnuceq},
\end{equation}
where $\Delta t_{\rm nuc}$ is the duration of the nucleation period   
\cite{IS,Ignatius}.                                                   

Surface tension and latent heat are provided by lattice
simulations with quenched QCD only, giving the values $\sigma = 0.015
T_c^3$, $l = 1.4 T_c^4$~\cite{Iwasaki}.
Scaling the latent heat for the physical QCD 
leads us to take $l = 3 T_c^4$.
%
Using              
these values 
the amount of
supercooling is $\Delta_{\rm sc} = 2.3 \times 10^{-4}$,
with large error bars.     
Due to rapid change of energy density near $T_c$, the microscopic sound  
speed in the quark phase has a small value,                              
$3 c_{\rm s}^2(T_{\rm c}) < 1$                                           
(for references see \cite{IS}).                                          

In the real Universe the local temperature of the radiation 
fluid fluctuates. 
The temperature contrast is denoted by $\Delta
\equiv \delta T/ \bar{T}$. On subhorizon scales in the radiation dominated 
epoch, each Fourier coefficient $\Delta(t,k)$ oscillates with constant 
amplitude
$\Delta_T(k)$. Inflation predicts a Gaussian 
distribution,
\begin{equation}
\label{pd}
p(\Delta){\rm d}\Delta = {1\over \sqrt{2\pi}\Delta_T^{\rm rms}}
\exp\left( - \frac12 {\Delta^2\over (\Delta_T^{\rm rms})^2}\right) 
{\rm d}\Delta \ .
\end{equation}
We find \cite{norm} for the COBE normalized \cite{Bennett} rms temperature 
fluctuation of the radiation fluid (not of cold dark matter)  
$\Delta_T^{\rm rms} = 1.0 \times 10^{-4}$ for a primordial Harrison-Zel'dovich
spectrum. The change of the equation of state prior to the QCD transition  
modifies the temperature-energy density relation, 
$\Delta = c_s^2 \delta\varepsilon /(\varepsilon + p)$. 
We may neglect the pressure $p$ 
near the critical temperature since 
$p \ll \varepsilon_{\rm q}$ at $T_{\rm c}$. 
On the other hand the drop of the sound speed enhances the amplitude of 
the density fluctuations proportional to $c_s^{-1/2}$ 
\cite{SSW}.             
Putting all
those effects together and allowing for a tilt in the power spectrum, 
the COBE normalized rms temperature fluctuation reads
\begin{equation}
\Delta_T^{\rm rms} \approx 10^{-4} (3 c_s^2)^{3/4} 
\left({k\over k_0}\right)^{(n-1)/2} \ .
\end{equation}
For a Harrison-Zel'dovich spectrum 
and
$3 c_s^2 = 0.1$, we find $\Delta_T^{\rm rms} \approx 2 \times 10^{-5}$. 

A small scale cut-off in the spectrum of primordial temperature fluctuations
comes from collisional damping by neutrinos 
\cite{SSW}.                      
For $l_{\rm smooth}$, the length over which temperature is constant,    
we take the value $10^{-4} d_{\rm H}$ \cite{IS}.                        
The temperature fluctuations are 
frozen with respect to the time scale of nucleations. 

Let us now investigate bubble nucleation in a Universe with spatially
inhomogeneous temperature distribution. 
Nucleation effectively takes place while the 
temperature drops by the tiny amount $\Delta_{\rm nuc}$. To determine the 
mechanism of nucleation, we compare 
$\Delta_{\rm nuc}$ with the rms 
temperature fluctuation $\Delta_T^{\rm rms}$:

1. If $\Delta_T^{\rm rms} < \Delta_{\rm nuc}$, the probability to
      nucleate a bubble at a given time is {\em homogeneous} in space. This is
      the case of homogeneous nucleation. 

2. If $\Delta_T^{\rm rms} > \Delta_{\rm nuc}$, the probability to
      nucleate a bubble at a given time is {\em inhomogeneous} in space. 
      We call this inhomogeneous nucleation.

\begin{figure}[t]
\begin{center}
\epsfig{figure=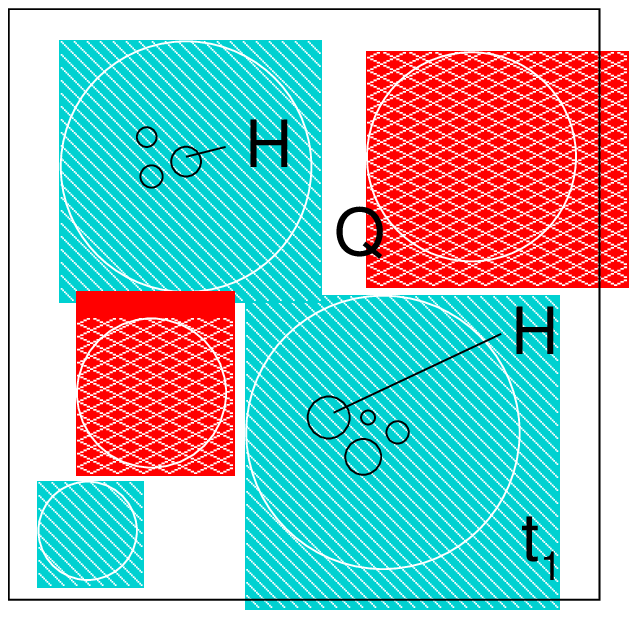,width=0.325\linewidth}          
\epsfig{figure=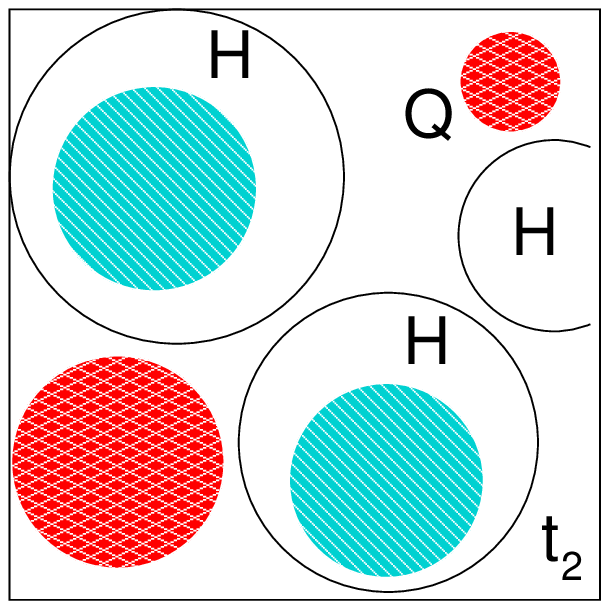,width=0.325\linewidth}          
\epsfig{figure=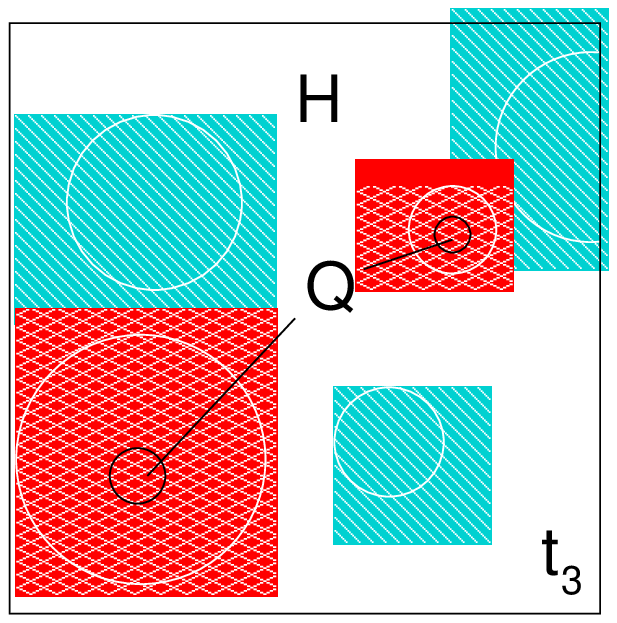,width=0.325\linewidth}          
\end{center}
\caption{\label{fig1}
Sketch of a first-order QCD transition in the inhomogeneous Universe. 
At $t_1$ the first hadronic bubbles (H) nucleate at the coldest spots
(light gray), while most of the Universe remains in the quark phase (Q). 
At $t_2$ the bubbles inside the cold spots have merged and have grown 
to bubbles as large as the temperature fluctuation scale. At $t_3$ the 
transition is almost finished. The last quark droplets are found in the 
hottest spots (dark gray).}
\end{figure}
The quenched lattice QCD data and a COBE normalized flat spectrum 
lead to 
the values $\Delta_{\rm nuc} \sim 10^{-6}$ and $\Delta_T^{\rm rms} \sim 
10^{-5}$. We conclude that the cosmological QCD transition may proceed
via inhomogeneous nucleation. A sketch of inhomogeneous nucleation is shown 
in Fig.~\ref{fig1}. The basic idea is that temperature inhomogeneities 
determine the location of bubble nucleation. Bubbles nucleate first in the 
cold regions.

For the fastest 
fluctuations, with angular frequency $c_s/l_{\rm smooth}$, we find
\begin{equation}
\label{coolingrate}
{{\rm d}T(t,{\bf x})\over {\rm d} t} = {\bar T\over t_{\rm H}}
\left[- 3 c_s^2 + 
{\cal O}\left(\Delta_T {t_{\rm H}\over \delta t}\right)\right]\ .
\end{equation}
The Hubble expansion is the dominant contribution, as typical values
are $3 c_s^2 = 0.1$ from quenched lattice QCD and $\Delta_T^{\rm rms}
t_{\rm H}/\delta t \approx 0.01$. 
This means that the local temperature does never increase, except by 
the released latent heat during bubble growth.

To gain some insight in the physics of inhomogeneous nucleation,
let us first inspect a simplified case. We have some randomly 
distributed cold spheres of diameter $l_{\rm smooth}$ with equal and uniform 
temperature, which is by the amount $\Delta_T^{\rm rms}T_c$ smaller than 
the again uniform temperature in the rest of the Universe. When the 
temperature in the cold spots has dropped to $T_{\rm f}$,
homogeneous nucleation takes place in them. Due to the
Hubble expansion the rest of the Universe would need the time
$\Delta t_{\rm cool} = t_H \Delta^{\rm rms}_T  / 3 c_s^2$          
to cool down to $T_{\rm f}$. 
The cold spots have fully 
been transformed into the hadron phase while the rest of the Universe 
still is in the quark phase. The latent heat released in a cold spot 
propagates in all directions, which provides the length scale
\begin{equation}
\label{lheat}
  l_{\rm heat} \equiv 2 v_{\rm heat} \Delta t_{\rm cool}. 
\end{equation}
If the typical distance from the boundary of a cold spot to the boundary
of a neighboring cold spot is less than 
$l_{\rm heat}$, then no hadronic bubbles can nucleate in the intervening 
space. 

The real Universe consists of smooth patches of typical linear size 
$l_{\rm smooth}$, their temperatures given by the distribution~(\ref{pd}).
At time $t$ heat, 
coming from a cold spot which was transformed into hadron phase at time 
$t'$, occupies the volume
$V(t,t') = (4\pi/3) [l_{\rm smooth}/2 + v_{\rm heat}(t-t')]^3$.
The fraction of space that is not reheated by the released latent heat 
(and not transformed to hadron phase), is given at time $t$ by
\begin{equation}
\label{fihn}
f(t) \approx 1 - \int_0^{t}\Gamma_{\rm ihn}(t') V(t,t') {\rm d}t',
\end{equation}
where we neglect overlap and merging of heat fronts. 
$\Gamma_{\rm ihn}$ is the volume 
fraction converted into the new phase, per physical time and volume as a 
function of the mean temperature $T=\bar{T}(t)$. 
We find                         
\begin{equation}
\label{gammai}
\Gamma_{\rm ihn} = 3 c_s^2{T_f\over T}{1\over t_{\rm H} {\cal V}_{\rm smooth}}
p(\Delta = \frac{T_{\rm f}}{T} - 1) , 
\end{equation}
where the relevant physical volume is ${\cal V}_{\rm smooth} = 
(4\pi/3) (l_{\rm smooth}/2)^3.$

The end of the nucleation period, $t_{\rm ihn}$, is defined through 
the condition $f(t_{\rm ihn}) = 0$. We 
introduce the variables $N \equiv (1 - T_{\rm f}/T)/\Delta_T^{\rm rms}$ and 
${\cal N} \equiv N(t_{\rm ihn})$. 
Putting everything together we determine 
${\cal N}$ from
\begin{equation}
{l_{\rm heat}^3\over l_{\rm smooth}^3} \int_{\cal N}^\infty {\rm d} N 
\frac{e^{-\frac12 N^2}}{\sqrt{2\pi}} 
\left(\frac{l_{\rm smooth}}{l_{\rm heat}} + N - {\cal N}\right)^3 = 1 .
\end{equation} 
For $l_{\rm heat}/l_{\rm smooth} = 1, 2, 5, 10$ we find 
${\cal N} \approx 0.8, 1.4, 2.1, 2.6$, respectively.

The effective nucleation distance in inhomogeneous nucleation is
defined from the number density of 
those cold spots that acted as nucleation centers, 
$d_{\rm nuc,ihn} \equiv n^{- 1/3}$. 
We find
\begin{eqnarray}
d_{\rm nuc,ihn} & \approx &
\left[\int_0^{t_{\rm ihn}} \Gamma_{\rm ihn}(t) {\rm d} t\right]^{-1/3} \\
& = & 
[\frac{3}{\pi} (1 - {\rm erf}({\cal N}/\sqrt{2})]^{-1/3} \, l_{\rm smooth}. 
\end{eqnarray}
With the above values $l_{\rm heat}/l_{\rm smooth} = 1, 2, 5, 10$ we get 
$ d_{\rm nuc,ihn} = 1.4, 1.8, 3.0,
$ $            
4.8 \times l_{\rm smooth}$, where
$l_{\rm smooth} \approx 1 \mbox{\ m}$. 

For a COBE normalized spectrum without any tilt and with 
a tilt of $n-1 = 0.2$,
together with $3c_s^2 = 0.1$ and $v_{\rm heat} = 0.1$, we find the estimate 
$l_{\rm heat}/l_{\rm smooth} \approx 0.4$ and 9, correspondingly. 
Notice that the values of $v_{\rm heat}$ and $3 c_s^2$ are 
in principle unknown.
Anyway, we can conclude that the case $\l_{\rm heat} > l_{\rm smooth}$
is a realistic possibility. 

With $2 v_{\rm heat}(3c_s^2)^{-1/4} (10^{-4}d_{\rm H}/l_{\rm smooth}) < 1$
and without positive tilt
we are in the region $l_{\rm heat} < l_{\rm smooth}$, where the geometry is 
more complicated and the above quantitative analysis does not apply.
In this situation nucleations take place in 
the most common cold spots (${\cal N} 
\sim 1$), which are very close to each other. We expect a structure of 
interconnected baryon-depleted and baryon-enriched layers with typical surface
$l_{\rm smooth}^2$ and thickness 
$l_{\rm def} \equiv v_{\rm def}\Delta t_{\rm cool}$. 

We emphasize that inhomogeneous and heterogeneous
nucleation~\cite{Christiansen} are genuinely different mechanisms, although 
they give the same typical scale of a few meters by chance. 
If latent heat and surface tension of QCD would turn out to reduce  
$\Delta_{\rm sc}$ to, e.g., $10^{-6}$, instead of $10^{-4}$,
the maximal heterogeneous nucleation distance would fall to the 
centimeter scale, whereas on the distance in 
inhomogeneous nucleation this would have no effect.

We have shown that inhomogeneous nucleation during the QCD transition 
can give rise to an inhomogeneity scale exceeding the proton diffusion
scale
(2~m at 150~MeV).                  
The resulting baryon inhomogeneities could provide inhomogeneous initial 
conditions for nucleosynthesis. Observable deviations from the element 
abundances predicted by homogeneous nucleosynthesis seem to be 
possible in that case~\cite{MathewsKainulainen}.

In conclusion, we found that inhomogeneous nucleation leads to 
nucleation distances that exceed by two orders of magnitude 
estimates based on homogeneous nucleation. We 
point out         
that this new 
effect appears for the (today) most probable range of cosmological and 
QCD parameters.

\section*{Acknowledgments}
We acknowledge Willy the Cowboy for valuable encouragement. We thank 
K.~Rummukainen for crucial help,
K.~Jedamzik for discussions,                 
and J.~Madsen for correspondence.
J.I.~would like to thank the Academy of Finland and
D.J.S.~the 
Austrian Academy of Sciences for financial support.
We are grateful to the organizers of the Strong and Electroweak Matter 
meetings; the research presented here was initiated at SEWM-97 in Eger, 
Hungary.

\end{document}